\documentstyle[12pt,epsfig]{article}
\newcommand{\be}{\begin{equation}}
\newcommand{\ee}{\end{equation}}
\newcommand{\bea}{\begin{eqnarray}}
\newcommand{\eea}{\end{eqnarray}}

\begin{document}

\title{
{\vspace{-1cm} \normalsize
\hfill \parbox{40mm}{         }}\\[25mm]
A New Gauge Fixing Method\\
 for Abelian Projection}

\author{
$^1$F.Shoji\thanks{shoji@riise.hiroshima-u.ac.jp}
,
$^2$T.Suzuki\thanks{suzuki@hep.s.kanazawa-u.ac.jp}
,
$^2$H.Kodama\thanks{h--kodama@hep.s.kanazawa-u.ac.jp}
 and
$^1$A.Nakamura\thanks{nakamura@riise.hiroshima-u.ac.jp}
\\
{\footnotesize $^1$ {\it Research Institute for 
Information Science and Education,}}\\
{\footnotesize {\it Hiroshima University, Kagamiyama, Higashi-Hiroshima,
739-8521, Japan}} \\
{\footnotesize $^2$ {\it Institute for Theoretical Physics,}}\\
{\footnotesize {\it Kanazawa University, Kakuma, Kanazawa, 920-1192, Japan}} 
}

\maketitle

\begin{center} 
\begin{bf}
Abstract
\end{bf}
\end{center}
We formulate a stochastic gauge fixing method to 
study the gauge dependence of the Abelian projection.
We consider a gauge which interpolates between the 
maximal Abelian gauge and no gauge fixing.
We have found that Abelian dominance for the heavy quark potential 
holds even in a gauge which is far from maximally Abelian one.
The heavy quark potentials from monopole and photon contribution 
are calculated at several values of the gauge parameter,
and the former part shows always the confinement behavior.

\section{Introduction}

Since 'tHooft and Mandelstam proposed the QCD vacuum state to behave
like a magnetic superconductor, a dual Meissner effect has been considered
to play an essential role in the mechanism of color confinement
\cite{thooft,mand}.
A gauge is chosen to reduce the gauge symmetry of a non-Abelian group
to its maximal Abelian (MA) subgroup, and Abelian fields and magnetic
monopole can be identified there.
When one reduces $SU(N)$ to $U(1)^{N-1}$ by 
the partial gauge fixing, 
monopoles appear in $U(1)^{N-1}$ sector as a topological object.
Confinement of QCD is conjectured to be due to  
condensation of the monopoles.
By using the MA gauge which maximizes the functional
\begin{eqnarray}
R&=&\sum_{x,\mu} \mbox{Tr}~ U_\mu(x)\sigma_3 U_\mu(x)^{\dagger}\sigma_3,
\end{eqnarray}
\noindent
Suzuki and Yotsuyanagi\cite{suzuki-yotsu} first found that the 
value of Abelian string
tension is close to that of the non-Abelian theory,
where $U_\mu(s)$ are $SU(2)$ link variables on the lattice.
Since then many numerical evidences have been collected to show
the importance of monopoles in QCD vacuum: we refer to Ref.\cite{Suzuki-Review}
for a review of these results.

There are infinite ways of extracting $U(1)^{N-1}$ from $SU(N)$. 
This corresponds to the choice of gauge in Abelian projection.
Abelian and monopole dominances can be clearly seen in MA gauge
but not in the others;
they seem to depend on the choice of gauge in the Abelian projection.
However, the dual Meissner effect only in MA gauge is not enough for the 
proof of color confinement, since Abelian charge confinement and color
confinement are different.

Recently Ogilvie\cite{ogilvie} has developed a character
expansion for Abelian and found that gauge fixing is unnecessary, i.e.,
Abelian projection yields string tensions
of the underlying non-Abelian theory even {\it without} gauge
fixing. Essentially the same mechanism was observed by Ambj{\o}rn and Greensite
for $Z_2$ center projection of $SU(2)$ link variables\cite{ambjorn}.
See also Ref.\cite{greensite}.
Furthermore, by introducing a gauge fixing function 
$S_{gf}=\lambda \sum \mathrm{Tr} U_\mu(x)\sigma_3
U_\mu(x)^\dagger\sigma_3$, 
Ogilvie has also shown that the Abelian dominance for
the string tension occurs for small $\lambda$.
Hence he conjectures that Abelian dominance is gauge independent and that 
gauge fixing results in producing fat links for 
Wilson loop and is computationally advantageous for the measurements.  
Further Suzuki et.al. have shown that if the gauge independence of Abelian
dominance is realized, the gauge independence of monopole dominance is also
proved\cite{suzuki}.
Hence to prove the gauge independence of Abelian and monopole dominances 
are very important especially in the intermediate region 
between no gauge fixing and exact MA gauge fixing. 

In this letter, we analyze the gauge dependence of the Abelian 
projection.
We now employ stochastic quantization with gauge fixing as 
the gauge fixing scheme which has been proposed by Zwanziger\cite{zwanziger}

\be
\frac{\partial}{\partial\tau}A^a_\mu(x,\tau)=
-\frac{\delta S}{\delta
A^a_\mu(x,\tau)} + \frac{1}{\alpha}D^{ab}_\mu \Delta^b 
 + \eta^a_\mu(x,\tau),
\ee
where $x$ is Euclidean space-time and $\tau$ is fictious time.
$\eta$ stands for Gaussian white noise 
\bea
\langle\eta^a_\mu(x,\tau)\rangle&=&0,\nonumber
\eea
\bea
\langle\eta^a_\mu(x,\tau)\eta^b_\nu(x',\tau')\rangle&=&
2\delta^{ab}\delta_{\mu\nu}\delta^4(x-x')\delta(\tau-\tau').\nonumber
\eea
Here $\Delta$ is defined as
\bea
\Delta(x)=\Delta^1(x)\sigma_1+\Delta^2(x)\sigma_2,\nonumber
\eea
\bea
\Delta^1(x)&=&-4g\sum_\mu[\partial_\mu
A^1_\mu(x)-g A^2_\mu(x)A^3_\mu(x) ],\\
\Delta^2(x)&=&-4g\sum_\mu[\partial_\mu A^2_\mu(x)+g A^1_\mu(x)A^3_\mu(x) ].\
\eea
\noindent
Note that
$\alpha=0$ corresponds to the MA gauge fixing and $\alpha=\infty$ is
the stochastic quantization without gauge fixing.

\section{Formulation}

$SU(2)$ elements can be decomposed into diagonal and off-diagonal parts after
Abelian projection, 
\begin{eqnarray}
U_\mu(x)&=&c_\mu(x)u_\mu(x),
\end{eqnarray}
where $c_\mu(x)$ is the off-diagonal part and $u_\mu(x)$ is the diagonal one,
\[ u_\mu(x) = \left(
	\begin{array}{@{\,}cc@{\,}}
	\exp(i \theta_\mu(x) ) & 0 \\
	0                      & \exp(-i \theta_\mu(x) ) 
	\end{array}
\right) . \]
\noindent
The diagonal part can be regarded as link variable of the remaining U(1).
One can construct monopole currents from 
field strength of U(1) links\cite{degrand}:
\begin{eqnarray}
\theta_{\mu\nu}(x)&=&\theta_\mu(x)+\theta_\nu(x+\hat{\mu})
-\theta_\mu(x+\hat{\nu})-\theta_\nu(x)\nonumber\\
&=&\bar{\theta}_{\mu\nu}+2\pi n_{\mu\nu}~~~
 (-\pi\leq\bar{\theta}_{\mu\nu}<\pi) , \nonumber\\
k_\mu(x)&=&\frac{1}{4\pi}\epsilon_{\mu\nu\rho\sigma}  
\partial_\nu \bar{\theta}_{\rho\sigma}(x) .
\end{eqnarray}

Wilson loops from Abelian, monopole and photon
contributions can be calculated as \cite{siba}
\bea
W^{{\rm Abelian}}&=&{\rm exp}(-\frac{i}{2}\sum_{x,\mu,\nu} M_{\mu\nu}(x)
\theta_{\mu\nu}(x)),\\
W^{{\rm monopole}}&=&{\rm exp}(2\pi i\sum_{x,x',\alpha,\beta,\rho,\sigma}
k_\beta(s)D(x-x') \frac{1}{2}\epsilon_{\alpha\beta\rho\sigma}
\partial_\alpha M_{\rho\sigma}(x')),\\
W^{{\rm photon}}&=&{\rm exp}(-i\sum_{x,x',\mu,\nu}
\partial_\mu^-\theta_{\mu\nu}(x)D(x-x')J_\nu(x')),\\
&&J_\nu(x)=\partial^-_\mu M_{\mu\nu}(x),\nonumber
\eea
where $\partial$ is a lattice forward derivative, $\partial^-$ is a backward
derivative and $D(x-x')$ is the lattice Coulomb propagator.
$J_\nu$ is the external source of electric charge and
$M_{\mu\nu}$ has values $\pm1$ on the surface inside of Wilson loop.
In order to achieve better signals,
we perform smearing for spatial link variables of non-Abelian, 
Abelian, monopole and photon\cite{bali}
.We set $\gamma=2.0$ for non-Abelian configurations 
and $\gamma=1.0$ for the others, where $\gamma$ is the parameter which 
determines the mixing between a link variable itself and staples
surrounding the link
\[
U_\mu(x)^{{\mathrm smeared}}=\frac{1}{N}\{\gamma~U_\mu(x) + \sum
({\mathrm staples})\}.
\]
Here $N$ is a normalization factor.

Stochastic quantization is based on Langevin equation which describes 
stochastic processes in terms of fictious time\cite{parisi}.
A compact lattice version of this equation with gauge fixing
was proposed in Ref.\cite{mizu}:
\bea
U_\mu(x,\tau+\delta\tau)&=&\omega(x,\tau)^{\dagger}
{\rm exp}(if^a_\mu \sigma_a)
U_\mu(x,\tau)\omega(x+\hat{\mu},\tau),\\
f_\mu^a&=&-\frac{\partial S}{\partial
A^a_\mu}\delta\tau+\eta^a_\mu(x,\tau)
\sqrt{\delta\tau},\nonumber\\
\omega(x,\tau)&=&{\rm exp}(i\frac{\beta}{2 N_c\alpha}
\Delta^a_{\mathrm lat}(x,\tau) 
\sigma_a \delta\tau).\nonumber
\eea
In MA gauge, 
\bea
\Delta_{\mathrm lat}(x,\tau)&=&i[\sigma_3,X(x,\tau)]\nonumber\\
&=&2(X_2(x,\tau)\sigma_1-X_1(x,\tau)\sigma_2),
\eea
where
\bea
X(x,\tau)&=&\sum_{\mu}(U_\mu(x,\tau)\sigma_3U_\mu(x,\tau)^\dagger
-U_\mu(x-\hat{\mu},\tau)^\dagger\sigma_3U_\mu(x-\hat{\mu},\tau))\\
&=&\sum_i X_i(x,\tau)\sigma_i.
\eea

As an improved action to reduce finite lattice spacing effects,
we adopt the Iwasaki action\cite{iwasaki}

\begin{figure}[h]
\vspace{-0.5cm}
\begin{center}
\psfig{file=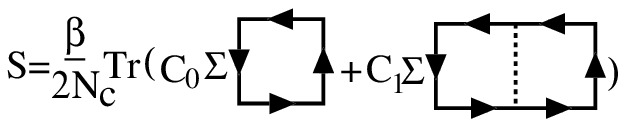,width=6cm}
\end{center}
\vspace{-0.2cm}
\end{figure}

\noindent
where $C_0+8C_1=1$ and $C_1=-0.331$.
The Runge--Kutta algorithm is employed for solving the 
discrete Langevin equation\cite{runge}. 
As will be shown, the systematic error
which comes from finite $\delta \tau$ is much reduced.

\section{Numerical Results}

Numerical simulations were performed on $8^3\times12$ and 
$16^3\times24$ lattices with
$\beta=0.995$, $\alpha=0.1,0.25,0.5,1.0$, and  $\delta\tau=0.001,0.005,0.01$.
Measurements were done every 100--1000
Langevin time steps 
after 5000--50000 thermalization Langevin time steps.
The numbers of Langevin time steps for the thermalization were
determined by monitoring the functional $R$ and Wilson loops.

\begin{figure}[htb]
\begin{center}
\psfig{file=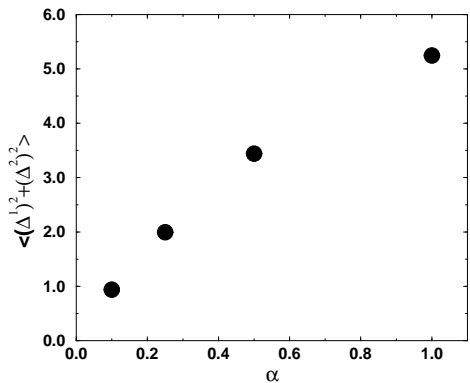,width=7cm}
\end{center}
\vspace{-1.cm}
\caption{
Gauge parameter $\alpha$ versus $(\Delta^1)^2+(\Delta^2)^2$.
Lattice size is $8^3\times12$, $\delta\tau=0.005$ and $\beta=0.995$.
}
\label{alphaDelta}
\vspace{-0.2cm}
\end{figure}

In Fig.\ref{alphaDelta}, we plot $(\Delta^1)^2+(\Delta^2)^2$ 
as a function of the
gauge parameter $\alpha$. 
$\alpha=0$ corresponds to $\Delta=0$, i.e., MA gauge.
When $\alpha$ increases, the deviation from the gauge fixed plane
becomes larger. 

\vspace{-.7cm}

We calculated the heavy quark potentials from non-Abelian, Abelian, monopole
and photon contributions by
\bea
V(R)=-\lim_{T\rightarrow\infty}\log\frac{\langle W(R,T)\rangle}
{\langle W(R,T-1)\rangle}.
\eea
We fit them with the following function,
\bea
V(R)=V_0+\sigma R + \frac{e}{R}.
\eea

\begin{figure}[htb]
\vspace{-0.2cm}
\begin{center}
\psfig{file=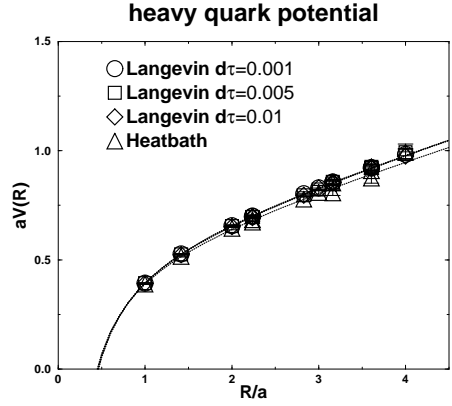,width=7cm}
\end{center}
\vspace{-1.cm}
\caption{
Non-Abelian heavy quark potentials derived from heatbath and 
Langevin updates for three values of $\tau$. 
Lattice size is $8^3\times 12$ and $\beta=0.995$.
}
\label{pot-nonabel}
\vspace{-0.2cm}
\end{figure}

In order to check that our Langevin update algorithm with the 
stochastic gauge fixing term works correctly, 
we plot in Fig.\ref{pot-nonabel} the heavy
quark potential $V(R)$, which is consistent with the result by 
the heatbath update.
Runge--Kutta method works well, i.e., 
we see no difference among data with $\delta \tau = 0.001, 0.005$ and $0.01.$ 

\begin{figure}[htb]
\vspace*{-0.2cm}
\begin{center}
\psfig{file=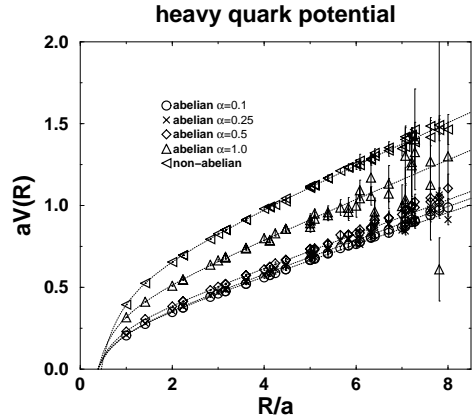,width=7cm}
\end{center}
\vspace{-1.cm}
\caption{
Heavy quark potentials from non-Abelian and Abelian contributions.
Lattice size is $16^3\times 24$, $\delta\tau=0.005$ and $\beta=0.995$.
}
\label{pot-abel}
\vspace{-0.2cm}
\end{figure}

\begin{figure}[htb]
\vspace*{-0.2cm}
\begin{center}
\psfig{file=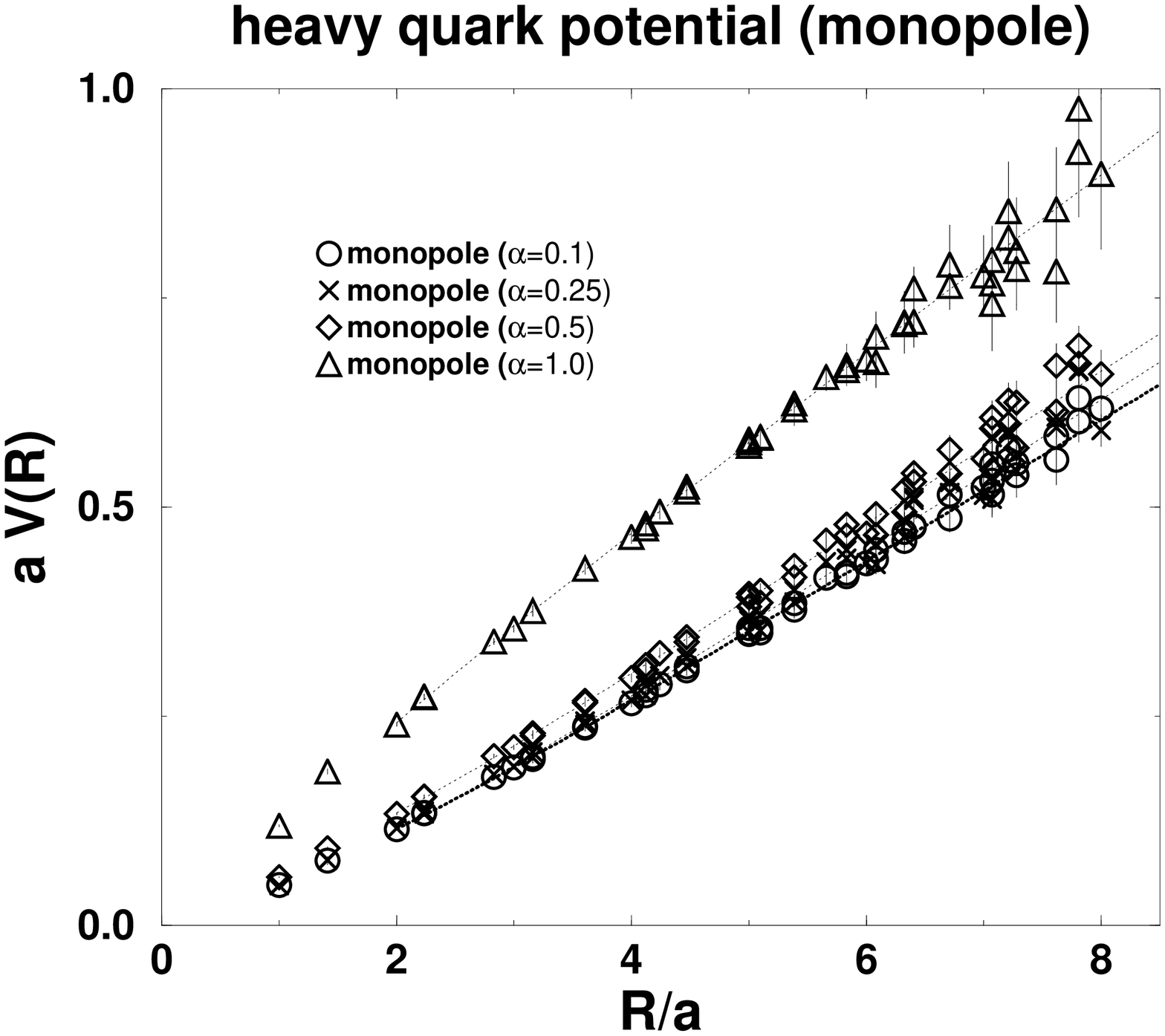,width=6.5cm}
\psfig{file=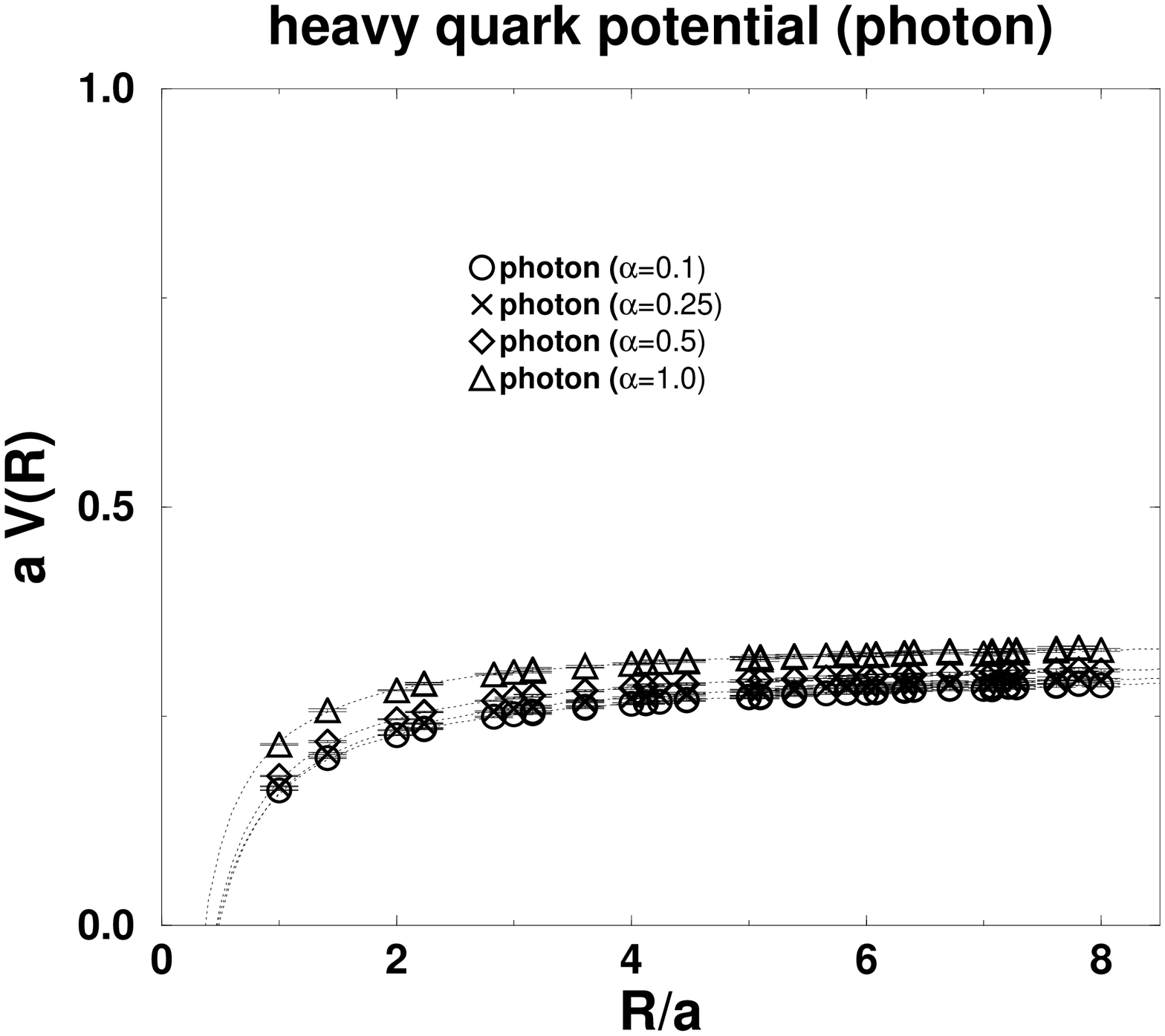,width=6.5cm}
\end{center}
\vspace{-1.cm} 
\caption{
Heavy quark potentials from monopole(left) and photon(right) 
contributions. Lattice size is $16^3\times24$, $\delta\tau=0.005$ and 
$\beta=0.995$.
}
\label{pot-mono}
\vspace{-0.2cm}
\end{figure}

In Fig.\ref{pot-abel} we show the Abelian heavy quark potentials 
for different $\alpha$'s
together with that of non-Abelian potential.
The heavy quark potentials from monopole and photon contributions are
plotted in Fig.\ref{pot-mono}.
They can be well fitted by a linear and Coulomb terms, respectively.
We see that the linear parts of potentials are essentially 
same from $\alpha=0.1$ to $1.0$,
and all of them show the confinement linear potential behavior.
Therefore even when we deviate from the MA gauge fixing condition,
we can identify the monopole contribution of the heavy quark potential
showing the confinement behavior.
As $\alpha$ increases, statistical error becomes larger.
This result suggests that the gauge fixing is favorable for decreasing 
numerical errors as pointed by Ogilvie\cite{ogilvie}.

\begin{figure}[htb]
\vspace*{-0.2cm}
\begin{center}
\psfig{file=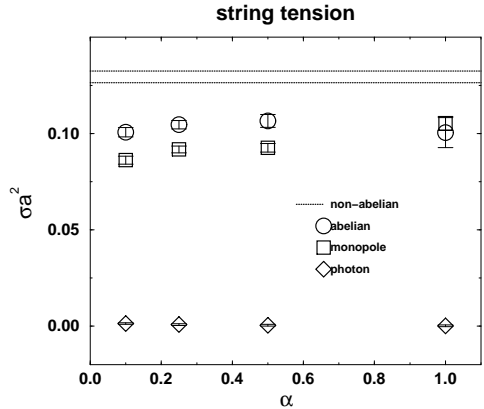,width=7cm}
\end{center}
\vspace{-1.cm}
\caption{
String tensions from non-Abelian, Abelian, monopole and photon
 contributions.
Lattice size is $16^3\times24$, $\delta\tau=0.005$ and $\beta=0.995$.}
\label{string}
\end{figure}

In Fig.\ref{string} we plot the values of the string tensions from Abelian,
monopole and photon contributions as a function of the gauge parameter
$\alpha$.
They are obtained by fitting the data in the range $2.0\le R\le 7.0$.
We have taken into account only statistical errors.
The upper two lines stand for the range of the non-Abelian string
tension.
The Abelian and the monopole dominances are observed for all values of
$\alpha$.
The string tensions from the Abelian parts are about 
80\% of the non--Abelian one. 
We expect that the difference of the percentage between our result and 
that of Bali et.al.\cite{bali} 
becomes smaller when we go to larger lattice size.
On the other hands, the string tension from the photon part is consistent
with zero.

\section{Concluding Remarks}

We have developed a stochastic gauge fixing method
which interpolates between the MA gauge and no gauge fixing.
In Refs.\cite{ogilvie},\cite{ambjorn} and \cite{greensite},
effects of gauge fixing are studied for lattice algorithms 
where gauge fixing is done after field configurations are generated.
In the stochastic gauge fixing procedure studied here,
the attractive force to the gauge fixed plane along a gauge orbit
is applied together with the Langevin update force. 
The method is tested together with the Iwasaki improved action
and the Runge-Kutta algorithm. We have found it works well.

We have studied the gauge dependence of Abelian projected heavy 
quark potential. 
It is observed that the confinement force is essentially independent of
the gauge parameter.
In the calculation of Abelian heavy quark potential, we have seen
that as gauge parameter $\alpha$ increases, 
the statistical error becomes larger.
This result suggests that the gauge fixing is favorable for increasing the
statistics as pointed by Ogilvie\cite{ogilvie}.
It is expected that as $\alpha$
increase, Abelian string tension would approach the non--Abelian one
\cite{ogilvie, greensite}.
Therefore it is important to see behavior of the string tension
as $\alpha$ becomes much larger than one.  
But data are more noisy for large $\alpha$ and 
we are planning to employ a noise reduction technique 
such as integral method\cite{ppr}
for obtaining statistically significant data.

It is desirable to study the gauge dependence (or independence) 
of other quantities, such as the monopole condensation, 
which may reveal the role of gauge fixing in the dual superconductor scenario.
Hioki et.al. have reported the correlation between monopole density and
Gribov copy in MA gauge fixing\cite{hioki}. 
Gribov copy effect for Abelian string tension has been studied by
Bali et.al.\cite{bali}.
For Landau gauge, Gribov copy may be avoided by Langevin update algorithm
together with the stochastic gauge fixing\cite{mizu}. 
It is, therefore,  interesting to investigate effects of Gribov copy in Abelian 
projection with the method.

\vspace{.3cm}
T.S and A.N. acknowledge financial support from JSPS Grand-in Aid for
Scientific Research(~(B) No.10440073 and No.11695029 for T.S.) and
(~(C)No.10640272 for A.N.).

\end{document}